\documentclass[singlecolumn,pra]{revtex4}

\usepackage{graphicx}
\usepackage{dcolumn}
\usepackage{bm}
\usepackage{amssymb,amsmath}

\newcommand{\beq}{\begin{equation}}
\newcommand{\eeq}{\end{equation}}
\newcommand{\beqa}{\begin{eqnarray}}
\newcommand{\eeqa}{\end{eqnarray}}
\newcommand{\beqar}{\begin{eqnarray*}}
\newcommand{\eeqar}{\end{eqnarray*}}

\begin{document}
\input epsf
\title{\bf \large Local toy-model theory with ontic correlated states
  of composite systems}

\author{Berry Groisman}
\affiliation{Centre for Quantum Information and Foundations, Centre for
Mathematical Sciences, University of Cambridge, Wilberforce Road,
Cambridge CB3 0WA, United Kingdom. }

\begin{abstract}
We propose a toy-model theory, that mimics various characteristic features of
quantum mechanics. Unlike the toy-models previously studied in the
literature, our toy-model allows for an observer to have a full
knowledge of a system's real (ontic) state. This is achieved by introducing domains
of disjointness, that is by allowing ontic states to be
``non-orthogonal". The observer can perform tests which allow her to
distinguish between the states in a single domain of disjointness, but
not between all ontic states at once. The consequence of this
assumption is that the ontic picture is extended to include joint states of
two or more systems. This effectively amounts to emergence of entanglement in
the model. We argue that these features, albeit being a ``non-classical'' element in the theory, support the
view that quantum-mechanical states are ontic states.
\end{abstract}

\maketitle


\section{Epistemic models vs Ontic models}
Recently, Local Toy Model Theories (LTMT), initially proposed by Hardy
\cite{Hardy} and Spekkens \cite{Spekkens}, have attracted considerable attention. Hardy
was motivated by making a case for non-cloning being logically
independent of non-locality. Spekkens used his model to support the
conjecture that quantum states are states of observer's incomplete
knowledge (epistemic states), rather than real states of affairs (ontic
states). Following the two original articles few modified models were
proposed \cite{Sanders_at_all, vanEnk} and the general framework of
those models was discussed in various works \cite{Friederich, Shahar, Pusey, Coecke_Edwards, Knee}. The
main endeavour of all above works was directed at advocating the epistemic view of
quantum states \cite{comment,modal_theory}.

This article makes a case for ontic view of quantum states. We
discuss an alternative local toy-model (i.e. model that admits local
hidden variable description), where joint states of correlated individual systems exhibit
entanglement (thereby reinforcing recent claims that entanglement is
logically independent of non-locality \cite{Vedral}).

Our aim is to construct a toy-model theory in which states of physical systems imitate quantum states, yet the model is not a restriction of quantum theory. One of the central features of quantum mechanics is the existence of non-commutative observables and restrictions on their simultaneous measurements (uncertainty principle). Our model must mimic these features.
Our main thesis will be that in
the process constructing toy-models with those properties we face two options.

First, we can
avoid introducing entangled states by postulating restrictions on our
knowledge of the system's state of affairs. This is the route taken by Spekkens. The basic premiss here is that underlying ontic reality of a system is exhausted by a number of mutually disjoin ontic states. In order to achieve non-commutativity and no-cloning, one is forced to introduce limits on the observer's access to this underlying ontic reality, thereby giving (inevitable) rise to a distinction between ontic states and epistemic states - states to which the observer had access to. The ontic reality of states is maintained here only on the level of individual systems. Correlated (joint) states of composite systems are
mixtures of ontic states and do not have an ontic status of their own.

Second, we allow the observer to have full access to ontic states. However, we must put additional restrictions on those states in order to mimic non-commutativity, complementarity and no-cloning. The reward is, however, the possibility to extend ontic description to joint states, which is effectively amounts to entanglement. Thus, if we allow full knowledge of the system, then entanglement will be the price to pay. The
latter option is a good indication that quantum states are ontic states.

This article is organized as follows. In Section \ref{sec:whatisontic} we set up the general framework for our model and introduce the concept of a domain of disjointness. Section \ref{sec:2-level} presents the explicit construction of the model starting with two domains of disjointness in Sub-section \ref{sec:2domains} followed by four domains of disjointness in Sub-section \ref{sec:4domains}. This is followed by discussion in Section \ref{discussion}.

\section{What is ontic - single system}\label{sec:whatisontic}

The main paradigm of the realist's approach to physical reality
constitutes the view that a physical system can be in a number of
real physical states. In classical physics all possible states of the
system are perfectly distinguishable by appropriate tests and
measurements. Quantum-mechanical systems exhibit non-orthogonality
and complementarity. If our aim is to construct a toy-model that
mimics quantum mechanics we must follow one of the following two strategies.

First strategy is based on\\

 {\it Premise I:
The basic premise is that
  underlying ontic reality of any physical system in isolation
  constitutes of the set of mutually exclusive (disjoint)
  states. Disjointness of states implies their
  distinguishability. Therefore, one needs to impose limitations on
  access to those properties, e.g. limit distinguishability by hand, introduce hidden variables, knowledge balance principle, etc.}\\

Despite the fact that {\it Premise I} is deeply rooted in our everyday
life experience and in classical physics, there is no compelling
reason for why this should be taken as a fundamental starting point.

The second strategy is based on the following premise which, albeit being less intuitive, is nevertheless equally valuable and legitimate.\\

{\it Premise II:
The underlying ontic reality of any physical system
  constitutes of a certain set of states, $|s)\in \{S\}$, which
  exhausts all possible internal configurations (degrees of freedom)
  of that system {\it when it is isolated}. The states are not
  necessarily disjoint. There are subsets of mutual disjointness
  (domains of disjointness) $D$, each containing only mutually disjoint
  states. Two states $|s_i)$ and $|s_j)$ are disjoint if and only if
  they belong to the same domain of disjointness.}\\

It is reasonable to assume that for every states there must be at
least one disjoin states, therefore each $D$ must contain at least 2
states. One could raise an immediate objection to Premise II on the
basis that the assumption of disjoint ontic states as un-natural. We
stress however, that despite the fact that it will be an unusual
assumption to make for many physical systems, there is no inherent logical contradiction in this assumption.

Each $D$ corresponds to a test set-up (at least one), $M_D$, which
allows for an observer to distinguish between the states in $D$ with
certainty. Such a measurement will not disturb the states in $D$. A
measurement will disturb the states when implemented on the system
when it is in a state from a different $D$ and will yield an inconclusive result.
We define as measurement update rule in this case: the state of the system after the measurement $M_D$ will be one of the
states from $D$ with equal probability. This randomness is inherent property of the system.

Consider two different domains of disjointness $D_x$ and $D_y$, where
the system is in a state $|s)_x\in D_x$, while measurement is
$M_y$. The measurement setting $M_y$ will correspond to a set of
disjoint states $D_y=\{|s)_y\}$. In order to provide systematic
description of the measurement process we need to provide a rule that
relates $|s)_x$ with $D_y$. Thus, we define a map  $c:|s_i),|s_j)\in
D_\alpha\mapsto |s_k)\in D_\beta$.The action of $c(|s_i),|s_j))$ is
consistent with the postulate that there is no ontic reality beyond
the states in $S$ due to closure of $S$ under $c$.


We will consider systems with equal number of states in each
$D$. $n$-level systems is a system with $n$ states in each domain.
Let us describe properties of $c$ on the example of $n=2$.

\section{Two-level systems}\label{sec:2-level}
\subsection{Two domains of disjointness}\label{sec:2domains}
Two-level systems correspond to the case of 2 states in each domain of
disjointness. First, consider two domains of disjointness
$D_x=\{|0)_x,|1)_x\}$ and $D_y=\{|0)_y,|1)_y\}$, where we have four ontic
states in total. Each domain is associated with a corresponding
two-outcome test, $M_x$ and $M_y$ respectively. The test $M_x$ distinguishes
between the states in $D_x$ with certainty. It yields deterministic outcome
$m_x=0;1$ if prior to the measurement the system was in the state
$|0)_x$ or $|1)_x$ respectively. Thus, the states in $D_x$ fully characterize
the test set-up (up to the choice of labels of the outcomes).

What happens if $M_x$ is performed when the systems is in one of the
states in $D_y$? In this model, it yields an
inconclusive result.  Thus, we define an overlap between the states
$|0)_x$ and $|0)_y$, say, as the probability to obtain outcome $m_x=0$ when
the state is $|0)_y$, $p(m_x=0/0_y)=1/2$.

The fact, that we can link $M_x$ with the states of $D_y$ implies that
we should be able to define a mathematical relation between $D_x$ and
$D_y$. We define it as follows. The states
 $|0)_x$ and $|1)_x$ can be combined in two different ways, which give
 $|0)_y$ and $|1)_y$ respectively and vice versa, i.e.

\begin{eqnarray}\label{map}
c_{+_1}(0_x,1_x)\mapsto0_y,~~~c_{+_1}(0_y,1_y)\mapsto0_x\\
c_{-_1}(0_x,1_x)\mapsto1_y,~~~c_{-_1}(0_y,1_y)\mapsto1_x
\end{eqnarray}

\noindent $c_{\pm_1}$ are maps or rules, which are analogous to coherent superpositions in quantum mechanics and they
form an {\it isomorphism} between $D_x$
and $D_y$. The role of $c_{\pm_1}$ is two-fold. First, they provide an
ontological link between $D_x$ and $D_y$. Second, they allow us to
give a quantitative prediction of a result of a test $M_i$ performed when a system
is in $D_j$.\\

\noindent {\bf Properties of $c_{\pm_1}$:} The two maps are
rules, according to which the states are combined. Nevertheless, we
demand that they obey certain mathematical properties.\\

\noindent {\it Property 1:} $c_{\pm_1}$ can be only defined on states from a
single $D$.\\


\noindent {\it Property 2: Coherent superpositions of a state with itself.}
\begin{equation}\label{prop:sameD}
\begin{split}
&c_{+_1}(s,s)\mapsto s~\\
&c_{-_1}(s,s)\mapsto\emptyset ~\\
&c_{\pm_1}(s,\emptyset)\mapsto s\\
&c_{\pm_1}(\emptyset,s)\mapsto(\pm)s\\
&c_{+_1}(s,-s)=c_{-_1}(s,s)\\
&c_{\pm_1}(\emptyset,\emptyset)\mapsto\emptyset,
\end{split}
\end{equation}
where $\emptyset$ is a null-state. A null-state is not an ontic state
of the system. As seen from the forth line each state has its additive inverse. Additive inverse of a state is not ontologically distinct from the state and is not observable. The model would appear as more attractive without additive inverse, however additive inverse is necessary to maintain consistency between the mathematical properties of $c_{\pm_1}$ and permutation transformations (which will be described late-on). The existence of the additive inverse is makes our model closely resembling quantum theory in terms if the mathematical structure. Notice also, that $c_{+_1}$ does not obey additivity in the
sense the arithmetic operation of addition does.\\

\noindent {\it Property 3: Associativity}

\begin{equation}\label{prop:ass}
\begin{split}
c_{+_1}(c_{+_1}(s,t),c_{-_1}(s,t))=c_{+_1}(c_{+_1}(s,s),c_{-_1}(t,t)),\\
c_{-_1}(c_{+_1}(s,t),c_{-_1}(s,t))=c_{+_1}(c_{+_1}(t,t),c_{-_1}(s,s)).
\end{split}\end{equation}

\noindent Combination of Properties (\ref{prop:sameD}) and (\ref{prop:ass})
provides closure of the set $S$ under $c_{\pm_1}$. Indeed, it yields

\begin{equation}\begin{split}
&c_{+_1}(c_{+_1}(s,t),c_{-_1}(s,t))=c_{+_1}(s,\emptyset)\mapsto s,\\
&c_{-_1}(c_{+_1}(s,t),c_{-_1}(s,t))=c_{+_1}(t,\emptyset)\mapsto t.
\end{split}\end{equation}

\noindent For example,
\begin{equation}
c_{+_1}\left( c_{+_1}\left (|0)_y,|1)_y\right ),c_{-_1}(|0)_y,|1)_y)\right )=c_{+_1}(|0)_y,|\emptyset))\mapsto |0)_y.
\end{equation}

Since $c_{\pm}$ obeys associativity and distributivity properties, it will be much more convenient to change the notation from $c_{\pm}(|s),|t))$ to $|s)\pm_1 |t)$. However, it should be stressed again that despite resembling arithmetic addition and subtraction these are not arithmetic operations.

For example, consider the combination
\begin{eqnarray}
c_{+_1}(|0_x),|1_x))\equiv |0_x)+_1|1_x)\mapsto |0_y).
\end{eqnarray}
Each of the two superposed states can be represented in turn as a superposition of states from $D_y$, i.e.
\begin{eqnarray}
\left [ |0_y)+_1|1_y)\right ] +_1 [|0_y)-_1|1_y)]\mapsto |0_x)+_1|1_x)\mapsto |0_y).
\end{eqnarray}

\noindent {\bf Joint states of two systems:}
We will now extend our description to states of two or more systems. Such a combined state must include at least 16 states, namely \\

\begin{center}
\begin{tabular}{|c|c|c|c|}
\hline
$D_{xx}$&$D_{xz}$&$D_{zx}$&$D_{zz}$\\
\hline
$|0)|0)$&$|0)|0)$&$|0)|0)$&$|0)|0)$\\[0.5ex]

$|0)|1)$&$|0)|1)$&$|0)|1)$&$|0)|1)$\\[0.5ex]

$|1)|0)$&$|1)|0)$&$|1)|0)$&$|1)|0)$\\[0.5ex]

$|1)|1)$&$|1)|1)$&$|1)|1)$&$|1)|1)$\\[0.5ex]
\hline
\end{tabular}
\end{center}

Here these 16 states are divided into four domains of disjointness, e.g. $D_{xx}=\{|0)_x|0)_x,|0)_x|1)_x,|1)_x|0)_x,|1)_x|1)_x\}$ and so on. We will exclude sets of the type $\{|0)_x|0)_x,|0)_x|1)_x,|1)_x|0)_z,|1)_x|1)_z\}$ from being classified as domains of disjointness. We postulate that for each two states in a single $D$ the local parts must be either identical or disjoint.

Since {\it any} two disjoint states can be superposed coherently (postulated), the full set of states will include superposition of states like $|0)_x|0)_x$ and $|1)_x|1)_x$, etc.
We will show that no new type of maps is required to superpose joint states. Consider a superposition $c^{(2)}(|0)_x|0)_x,|0)_x|1)_x)$. It is obvious, that only the states of the second system are being actually superimposed while the state of the first system can be factored out, i.e.
\begin{equation}
c^{(2)}_{+_1}(|0)_x|0)_x,|0)_x|1)_x)=|0)_x c^{(1)}_{+_1}(|0)_x,|1)_x)=|0)_x [|0)_x+_1|1)_x]\mapsto |0)_x|0)_y.
\end{equation}
Thus, the map for 2 systems is of the same type as a map for one system and exhibits {\it distributivity} property. (We will therefore drop the superscript above the map.) In addition, by associativity (\ref{associativity}) we obtain a consistent result
\begin{equation}\begin{split}
|0)_x|0)_x +_1 |0)_x|1)_x &=|0)_x \left[|0)_y+_1 |1)_y\right ]+_1 |0)_x \left[|0)_z -_1 |1)_y\right]\\
&= \left[|0)_x|0)_y+_1 |0)_x|1)_y\right ]+_1 \left[|0)_x|0)_y -_1 |0)_x|1)_y\right]\\
&= \left[|0)_x|0)_y+_1 |0)_x|0)_y\right ]+_1 \left[|0)_x|1)_y -_1 |0)_x|1)_y\right]\\
&=|0)_x |0)_y +_1\emptyset=|0)_x |0)_y.
\end{split}\end{equation}

We arrive now at the main issue. Armed with the laws $c_{\pm}$ satisfy we consider superpositions of the states $|0)_x|0)_x$ and $|1)_x|1)_x$:
\begin{equation}\label{0x0x}
\begin{split}
|0)_x|0)_x&=[|0)_y+_1|1))_y][|0)_y +_1|1)_y]\\
&=|0)_y[|0)_y+_1|1)_y]+_1|1)_y [|0)_y+_1|1)_y]\\
&=[|0)_y|0)_y+_1|0)_y|1)_y]+_1[|1)_y|0)_y+_1|1)_y|1)_y]\\
&=[|0)_y|0)_y+_1|1)_y|0)_y]+_1[|0)_y|1)_y+_1|1)_y|1)_y],
\end{split}\end{equation}

\noindent where the second and third equalities are due to distributivity of $c_{\pm_1}$, while the last equality is due to associativity.
Similarly,
\begin{equation}\label{1x1x}
\begin{split}
|1)_x|1)_x&=[|0)_y-_1|1))_y][|0)_y -_1|1)_y]\\
&=|0)_y[|0)_y-_1|1)_y]-_1|1)_y [|0)_y-_1|1)_y]\\
&=[|0)_y|0)_y-_1|0)_y|1)_y]-_1[|1)_y|0)_y-_1|1)_y|1)_y]\\
&=[|0)_y|0)_y-_1|1)_y|0)_y]-_1[|0)_y|1)_y-_1|1)_y|1)_y],
\end{split}\end{equation}
which yields
\begin{equation}
|0)_x|0)_x +_1 |1)_x|1)_x=|0)_y|0)_y+_1|1)_y|1)_y.
\end{equation}

%

Similarly we can calculate other superpositions.
Thus, we obtain the following domain of disjointness $D_{ii}^c$ for correlated states, which are obtained by applying maps $c_{\pm_1}$ to product states of individual systems.
\begin{equation}\label{Dii}
\begin{split}
|0)_x|0)_x +_1 |1)_x|1)_x=&|0)_y|0)_y+_1|1)_y|1)_y)\\
|0)_x|0)_x -_1 |1)_x|1)_x=&|0)_y|1)_y+_1|1)_y|0)_y)\\
|0)_x|1)_x +_1 |1)_x|0)_x=&|0)_y|0)_y-_1|1)_y|1)_y)\\
|0)_x|0)_x -_1 |1)_x|0)_x=&|0)_y|1)_y-_1|1)_y|0)_y)\\
\end{split}
\end{equation}
In this domain the states exhibit definite parity with respect to tests performed on the subsystems in the same ``basis". These states are {\it ontic} states of a combined system. Thus, in our model correlated states of a composite system have ontic status of their own.

The second domain of disjointness for correlated states $D_{ij}^c$ corresponds to definite parity with respect to the tests performed in the complementary ``bases".
\begin{equation}
\begin{split}
|0)_x|0)_y +_1 |1)_x|1)_y=&|0)_y|0)_x+_1|1)_y|1)_x)\\
|0)_x|0)_y -_1 |1)_x|1)_y=&|0)_y|1)_x+_1|1)_y|0)_x)\\
|0)_x|1)_y +_1 |1)_x|0)_y=&|0)_y|0)_x-_1|1)_y|1)_x)\\
|0)_x|0)_y -_1 |1)_x|0)_y=&|0)_y|1)_x-_1|1)_y|0)_x)\\
\end{split}
\end{equation}

To emphasize the point that coherent superpositions are ontic states on their own right and shorten the notation we introduce the following notation $C_{\pm_1}^{ab}\equiv|0)_a|0)_b \pm_1 |1)_a|1)_b$ and $A_{\pm_1}^{ab}\equiv|0)_a|1)_b \pm_1 |1)_a|0)_b$ (for correlation and anti-correlation respectively).
The complete set of ontic states, divided into corresponding domains of disjointness is summarized in the following table.\\


\begin{center}
\begin{tabular}{|c|c|c||c|c|c|}
\hline
$D_{xx}$&$D_{yy}$&$D_{ab}^{c}$&$D_{xy}$&$D_{yx}$&$D_{ab}^{c}$\\[0.5ex]
\hline
~$|0)_x|0)_x$&$|0)_y|0)_y$&$C_{+_1}^{xx}=C_{+_1}^{yy}$&$|0)_x |0)_y$&$|0)_y|0)_x$&$C_{+_1}^{xy}=C_{+_1}^{yx}$\\[0.5ex]
~$|0)_x|1)_x$~&$|0)_y|1)_y$&$C_{-_1}^{xx}=A_{+_1}^{yy}$&$|0)_x|1)_y$&$|0)_y|1)_x$&$C_{-_1}^{xy}=A_{+_1}^{yx}$\\[0.5ex]
~$|1)_x|0)_x$~&$|1)_y|0)_y$&$A_{+_1}^{xx}=C_{-_1}^{yy}$&$|1)_x|0)_y$&$|1)_y|0)_x$&$A_{+_1}^{xy}=C_{-_1}^{yx}$\\[0.5ex]
~$|1)_x|1)_x$~&$|1)_y|1)_y$&$A_{-_1}^{xx}=A_{-_1}^{yy}$&$|1)_x|1)_y$&$|1)_y|1)_x$&$A_{-_1}^{xy}=A_{-_1}^{yx}$\\[0.5ex]
\hline
\end{tabular}
\end{center}

The states in each $D^c$ determine two {\it global} tests, $M_{ii}^c$ and $M_{ij}^c$, which distinguish between the states within $D_{ii}^{c}$ and
$D_{ij}^{c}$ respectively. The two test complement the {\it local} tests, $M_x$ and $M_y$.\\

{\bf Transformations:} For a single system we define 2 types of transformations. First, permutations within a single $D$, $P_x$ and $P_y$. Such permutations
have an affect on the complementary $D$, namely the state $|1)$ acquires a minus sign (becomes its own additive inverse), e.g. $P_x |1)_y=-|1)_y$. Second, a transformation between $D_x$ and $D_y$, $T_{x\leftrightarrow y}$, which is analogous to a rotation of spin in quantum theory.  \\

{\bf Dense Coding:} This simple model with only two domains of disjointness exhibits the phenomenon of dense coding in a manner completely analogous to QM. This should not surprise us, however, since so far our model looks very similar to a restricted version of quantum mechanics, if it is restricted to $x-z$ plane of the Bloch-sphere with only superpositions of equal amplitudes. The number of domains of disjointness for a single system is not a determining factor, it appears. What is important is the dimensionality of each domain (2), which leads to four (4) correlated coherent states in a $D^c$ and existence of 4 transformations on a single system, which map one correlated state to a itself or the other three. This is exactly how it works in our model.

Consider a state $C_{+_1}^{xx}$ shared between Alice and Bob. The four transformations Alice can perform are $I$, $P_x$, $P_y$ and $P_yP_x$.
\begin{center}
\begin{tabular}{rlrl}
$I$&$C_{+_1}^{xx}=$&$I$ &$C_{+_1}^{yy}\mapsto C_{+_1}^{xx}=C_{+_1}^{yy}$\\[0.5ex]
$P_x$&$C_{+_1}^{xx}=$&$P_x$ &$C_{+_1}^{yy}\mapsto A_{+_1}^{xx}=C_{-_1}^{yy}$\\[0.5ex]
$P_y$&$C_{+_1}^{xx}=$&$P_y$ &$C_{+_1}^{yy}\mapsto C_{-_1}^{xx}=A_{+_1}^{yy}$\\[0.5ex]
$P_y P_x$&$C_{+_1}^{xx}=$&$P_y P_x$ &$C_{+_1}^{yy}\mapsto A_{-_1}^{xx}=A_{-_1}^{yy}$
\end{tabular}
\end{center}

{\bf Teleportation:} Consider three systems $A_1$, $A_2$ and B, where systems $A_2$ and $B$ are in a correlated state $C_{+_1}^{xx}$. System $A_1$ is in an unknown arbitrary state, i.e. in one of the four ontic states. Consider all possible situations

\begin{equation}\label{teleporation}
\begin{split}
|0)_x C_{+_1}^{xx}=&~~~~~~~~~~~~~~~~~~~~~~~~~~~~~~~~~~~~~~~~~~~~~~~~~~~~~~~~~~~~~~~~~=(C_{+_1}^{xx}+_1C_{-_1}^{xx})|0)_x+_1(A_{+_1}^{xx}+_1A_{-_1}^{xx})|1)_x\\
|1)_x C_{+_1}^{xx}=&~~~~~~~~~~~~~~~~~~~~~~~~~~~~~~~~~~~~~~~~~~~~~~~~~~~~~~~~~~~~~~~~~=(A_{+_1}^{xx}-_1A_{-_1}^{xx})|0)_x+_1(C_{+_1}^{xx}-_1C_{-_1}^{xx})|1)_x\\
|0)_y C_{+_1}^{xx}=&|0)_y C_{+_1}^{yy}=(C_{+_1}^{yy}+_1C_{-_1}^{yy})|0)_y+_1(A_{+_1}^{yy}+_1A_{-_1}^{yy})|1)_y=(C_{+_1}^{xx}+_1A_{+_1}^{xx})|0)_y+_1(C_{-_1}^{xx}+_1A_{-_1}^{xx})|1)_y\\
|1)_y C_{+_1}^{xx}=&|1)_y C_{+_1}^{yy}=(A_{+_1}^{yy}-_1A_{-_1}^{yy})|0)_y+_1(C_{+_1}^{yy}-_1C_{-_1}^{yy})|1)_y=(C_{-_1}^{xx}-_1A_{-_1}^{xx})|0)_y+_1(C_{+_1}^{xx}-_1A_{+_1}^{xx})|1)_y,
\end{split}
\end{equation}
where we used the relation between correlated states written in
difference local bases in Eq. (\ref{Dii}).

It is clear, that measurement $M_{ii}^c$, which is analogous to Bell-measurement in quantum theory, will allow the parties to recover the original states at $B$ by applying corresponding corrective transformations on $B$ as follows
\begin{equation}
\begin{split}
&C_{+_1}^{xx}\rightarrow I \\
&C_{-_1}^{xx}\rightarrow P_y\\
&A_{+_1}^{xx}\rightarrow P_x\\
&A_{-_1}^{xx}\rightarrow P_yP_x,
\end{split}
\end{equation}
where $P_x$ and $P_y$ are permutations in $D_x$ and $D_y$ respectively. Thus, Alice sends 2 bits to Bob, who performs the corresponding correction.\\

\noindent It is tempting to associate a ``ball in a box"
interpretation of the ontic states, similar to one adopted by
Spekkens. Although, it is possible to do for a single system in our
model, such an interpretation cannot be extended to join states due to
their ontic nature. Thus, imposing such an interpretation even on a
single system would be misleading. The four ontic states
$\{|0)_x,|1)_x, |0)_y,|1)_y\}$ exhaust the ontological baggage of the
system. There is no more to it. After the observer performs one of the
tests $M_i$, there is nothing she is ignorant about - she has a full
knowledge about the system's state.

One might argue that this model is merely a restriction of quantum theory. This is true. However, we notice the following. First, our main motivation was to construct a model starting with a strong ontic premiss. If it leads to a model which resembles quantum theory, it is already an indication in favour of the ontic view of quantum states. Second, this model serves as an introduction to more sophisticated version which follows.

To summarize, We have considered a simple model with two domains of
disjointness. The model has a simple structure: ontic states with
coherent maps that relate between different domains of
disjointness. There is no complex vector space, no coefficients or
weights with which ontic states are superposed. We have shown how the
status of an ontic state can be given to a correlated state of two
systems.

\subsection{Four domains of disjointness}\label{sec:4domains}
To achieve a closer analogy with quantum mechanics additional domains
of disjointness are required. The natural step would be to add one
additional domain, $D_z$. This additional domain will require an
additional type of coherent maps, $c_{\pm_2}$. Thus, for example
\begin{equation}\begin{split}
|0)_x +_1 |1)_x \mapsto|0)_y,~~~~~|0)_x +_2 |1)_x\mapsto|0)_z\\
|0)_x -_1 |1)_x \mapsto|1)_y,~~~~~|0)_x -_2 |1)_x\mapsto|1)_z
\end{split}\end{equation}

\begin{figure} \epsfxsize=1.2truein
      \centerline{\epsffile{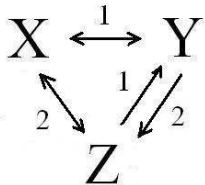}}
        \caption[]{Two different types of coherent maps
          (superpositions) connect the three domains of disjointness.}
    \label{3domains} \end{figure}

\noindent However, if one is trying to pursue this she will soon encounter a basic difficulty - the arrangement becomes frustrated. For example,

\begin{equation}\begin{split}
&|0)_y +_2 |1)_y \mapsto|0)_z,~~~~~|0)_z +_2 |1)_z=(|0)_y +_2 |1)_y)+_2 (|0)_y -_2 |1)_y)\mapsto|0)_y\\
&|0)_y -_2 |1)_y \mapsto|1)_z,~~~~~|0)_z -_2 |1)_z=(|0)_y +_2 |1)_y)-_2 (|0)_y -_2 |1)_y)\mapsto|1)_y,
\end{split}\end{equation}
which contradicts Fig. \ref{3domains}. In quantum theory this frustration is resolved by superposing states with complex coefficients. Here we aim to avoid complex vector spaces.

\begin{figure} \epsfxsize=1.2truein
      \centerline{\epsffile{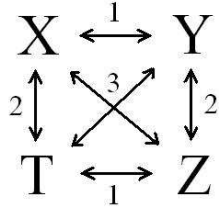}}
        \caption[]{Three different types of coherent maps
          (superpositions) connect the four domains of disjointness.}
    \label{4domains} \end{figure}

To overcome this difficulty, we will consider four domains of
disjointness. Despite the fact, that it takes us away from a direct
analogy with QM, it is important to push the limits of the model
without engaging with complex amplitudes. We introduce three different
types of coherent maps, each connecting a particular $D$ with three
other domains as shown below and summarized in Figure \ref{4domains}.

\begin{equation}
\begin{split}
&|0)_x +_1 |1)_x \mapsto|0)_y,~~~~~|0)_y+_2|1)_y\mapsto|0)_z,~~~~~|0)_z+_1|1)_z\mapsto|0)_t,~~~~~|0)_t+_2|1)_t\mapsto|0)_x,\\
&|0)_x -_1 |1)_x \mapsto|1)_y,~~~~~|0)_y-_2|1)_y\mapsto|1)_z,~~~~~|0)_z-_1|1)_z\mapsto|1)_t,~~~~~|0)_t-_2|1)_t\mapsto|1)_x,\\
\\
&|0)_y +_1 |1)_y\mapsto|0)_x,~~~~~|0)_z+_2|1)_z\mapsto|0)_y,~~~~~|0)_t+_1|1)_t\mapsto|0)_z,~~~~~|0)_x+_2|1)_x\mapsto|0)_t,\\
&|0)_y -_1 |1)_y\mapsto|1)_x,~~~~~|0)_z-_2|1)_z\mapsto|1)_y,~~~~~|0)_t-_1|1)_t\mapsto|1)_z,~~~~~|0)_x-_2|1)_x\mapsto|0)_t,\\
\\
&|0)_x +_3 |1)_x\mapsto|0)_x,~~~~~|0)_z+_3|1)_z\mapsto|0)_x,~~~~~|0)_y+_3|1)_y\mapsto|0)_t,~~~~~|0)_t+_3|1)_t\mapsto|0)_y,\\
&|0)_x -_3 |1)_x\mapsto|1)_x,~~~~~|0)_z-_3|1)_z\mapsto|1)_x,~~~~~|0)_y-_3|1)_y\mapsto|1)_t,~~~~~|0)_t-_3|1)_t\mapsto|0)_y,
\end{split}
\end{equation}

\noindent We impose the following properties on $c_{\pm_1}$, $c_{\pm_2}$ and $c_{\pm_3}$.\\

\noindent{\bf Property 1: Coherent superpositions of states of a single system.}
\begin{itemize}

\item Maps of the same type combined in the ``cascade" way
  satisfy distributivity and associativity properties as the maps in Section \ref{sec:2domains}. However,
  $\pm_3$ is not commutative, as will be specified below. Thus, for example,
\begin{equation}
|0)_x +_1 |1)_x = [|0)_y +_1 |1)_y]+_1 [|0)_y -_1 |1)_y] \mapsto|0)_y.
\end{equation}

\item Maps of different types combined obey
\begin{equation}
\begin{split}
[|0)_i +_2 |1)_i]+_1 [|0)_i -_2 |1)_i] = [|0)_i +_1 |1)_i]+_2 [|0)_i -_1 |1)_i=|0)_i+_3 |1)_i\mapsto |0)_j,\\
[|0)_i +_2 |1)_i]-_1 [|0)_i -_2 |1)_i] = [|0)_i +_1 |1)_i]-_2 [|0)_i -_1 |1)_i=|0)_i+_3 |1)_i\mapsto |1)_j,
\end{split}
\end{equation}
where $j$ is on the diagonal from $i$ in Fig. \ref{4domains}.

\item Special properties of $\pm_3$:
\begin{equation}\label{specialprop_3}
\begin{split}
|0)_i+_3 |1)_i=|1)_i-_3 |0)_i,\\
|0)_i-_3 |1)_i=|1)_i+_3 |0)_i.
\end{split}
\end{equation}

The map $\pm_3$ does not obey commutativity with respect to interchange of the arguments. These unusual properties are designed to satisfy the transformation rules, which will be discussed in the following.

\end{itemize}

\noindent{\bf Property 2: Coherent superpositions of joint states.}

\begin{itemize}

\item Coherent superpositions involving only $\pm_1$ for join states for all $D_{ii}$ with $i=x,y,z,t$ obey
distributivity and associativity as in Section \ref{sec:2domains}.

\item For all $D_{ii}$ ($i=x,y,z,t$) coherent superpositions where $\pm_1$, $\pm_2$ and $\pm_3$ are mixed obey

\begin{equation}
\begin{split}
[|0)_i +_2 |1)_i][|0)_i +_2 |1)_i]+_1 [|0)_i -_2 |1)_i][|0)_i -_2 |1)_i] = |0)_i|0)_i+_2 |1)_i|1)_i,\\
[|0)_i +_2 |1)_i][|0)_i +_2 |1)_i]-_1 [|0)_i -_2 |1)_i][|0)_i -_2 |1)_i] = |0)_i|1)_i+_2 |1)_i|0)_i,\\
[|0)_i +_2 |1)_i][|0)_i -_2 |1)_i]+_1 [|0)_i -_2 |1)_i][|0)_i +_2 |1)_i] = |0)_i|0)_i-_2 |1)_i|1)_i,\\
[|0)_i +_2 |1)_i][|0)_i -_2 |1)_i]-_1 [|0)_i -_2 |1)_i][|0)_i +_2 |1)_i] = |0)_i|1)_i-_2 |1)_i|0)_i,\\
\end{split}
\end{equation}

\begin{equation}
\begin{split}
[|0)_i +_1 |1)_i][|0)_i +_1 |1)_i]+_2 [|0)_i -_1 |1)_i][|0)_i -_1 |1)_i] = |0)_i|0)_i+_2 |1)_i|1)_i,\\
[|0)_i +_1 |1)_i][|0)_i +_1 |1)_i]-_2 [|0)_i -_1 |1)_i][|0)_i -_1 |1)_i] = |0)_i|1)_i+_2 |1)_i|0)_i,\\
[|0)_i +_1 |1)_i][|0)_i -_1 |1)_i]+_2 [|0)_i -_1 |1)_i][|0)_i +_1 |1)_i] = |0)_i|0)_i-_2 |1)_i|1)_i,\\
[|0)_i +_1 |1)_i][|0)_i -_1 |1)_i]-_2 [|0)_i -_1 |1)_i][|0)_i +_1 |1)_i] = |0)_i|1)_i-_2 |1)_i|0)_i,\\
\end{split}
\end{equation}

\begin{equation}
\begin{split}
[|0)_i +_2 |1)_i][|0)_i +_2 |1)_i]+_2 [|0)_i -_2 |1)_i][|0)_i -_2 |1)_i] = |0)_i|0)_i+_1 |1)_i|1)_i,\\
[|0)_i +_2 |1)_i][|0)_i +_2 |1)_i]-_2 [|0)_i -_2 |1)_i][|0)_i -_2 |1)_i] = |0)_i|1)_i+_1 |1)_i|0)_i,\\
[|0)_i +_2 |1)_i][|0)_i -_2 |1)_i]+_2 [|0)_i -_2 |1)_i][|0)_i +_2 |1)_i] = |0)_i|0)_i-_1 |1)_i|1)_i,\\
[|0)_i +_2 |1)_i][|0)_i -_2 |1)_i]-_2 [|0)_i -_2 |1)_i][|0)_i +_2 |1)_i] = |0)_i|1)_i-_1 |1)_i|0)_i,\\
\end{split}
\end{equation}

\begin{equation}\label{rule312}
\begin{split}
[|0)_i +_3 |1)_i][|0)_i +_3 |1)_i]+_1 [|0)_i -_3 |1)_i][|0)_i -_3 |1)_i] = |0)_i|0)_i+_2 |1)_i|1)_i,\\
[|0)_i +_3 |1)_i][|0)_i +_3 |1)_i]-_1 [|0)_i -_3 |1)_i][|0)_i -_3 |1)_i] = |0)_i|0)_i-_2 |1)_i|1)_i,\\
[|0)_i +_3 |1)_i][|0)_i -_3 |1)_i]+_1 [|0)_i -_3 |1)_i][|0)_i +_3 |1)_i] = |0)_i|1)_i+_2 |1)_i|0)_i,\\
[|0)_i +_3 |1)_i][|0)_i -_3 |1)_i]-_1 [|0)_i -_3 |1)_i][|0)_i +_3 |1)_i] = |0)_i|1)_i-_2 |1)_i|0)_i,
\end{split}
\end{equation}

with same rule as (\ref{rule312}) when $\pm_1$ and $\pm_2$ are interchanged.

\end{itemize}

\noindent Properties 1 and 2 provide consistent representation of correlated states in different local ``bases".
Using notation introduced in Section \ref{sec:2domains} we can write the first correlated domain of disjointness:

\begin{equation}\label{4Dii}
\begin{split}
&C_{+_1}^{xx}=C_{+_1}^{yy}=C_{+_2}^{zz}=C_{+_2}^{tt}\\
&C_{-_1}^{xx}=A_{+_1}^{yy}=C_{-_2}^{zz}=A_{+_2}^{tt}\\
&A_{+_1}^{xx}=C_{-_1}^{yy}=A_{+_2}^{zz}=C_{+_2}^{tt}\\
&A_{-_1}^{xx}=A_{-_1}^{yy}=A_{-_2}^{zz}=A_{-_2}^{tt}\\
\end{split}
\end{equation}

Thus, introducing the forth domain of disjointness allows our model to exhibit properties of correlations which are quite different from those in quantum theory, namely for a single joint state measured in different bases the toy-model exhibits even number of both correlations and anti-correlations. Unlike quantum mechanics, where for three domains of disjointness the number of anti-correlations is always odd, in this model it can be either even or odd. Thus we conjecture, that our model is not a straightforward restriction of quantum theory. The interesting question to be addressed is whether such correlations can be mimicked within quantum theory.\\

{\bf Transformations:} In the case of four domains of disjointness we define 3 types of transformations for a single system. First, $P_{xz}$ and $P_{yt}$ permute the states in two domains. When acting on the complementary domains these transformations cause $|1)$ to acquire a minus sign, e.g. $P_{xz} |1)_y=-|1)_y$. It is impossible to permute states in one domain of disjointness only.

Now it is clear that the properties (\ref{specialprop_3}) are the consequence of introduction of $P_{xz}$ and $P_{yt}$. \\

{\bf Dense Coding:} Dense coding works in a similar way to two domains of disjointness. Consider the state $C_{+_1}^{xx}$ shared between Alice and Bob. It is easy to check using  the properties of these state, that Alice can encode four bits of information by acting locally on her system with one of the  transformations $I$, $P_{xz}$, $P_{yt}$, or $P_{yt}P_{xz}$.\\

{\bf Teleportation:} Consider three systems $A_1$, $A_2$ and B, where systems $A_2$ and B are in a correlated state $C_{+_1}^{xx}$. System $A_1$ is in an unknown arbitrary state, i.e. in one of the six possible ontic states. Consider all possible situations

\begin{equation}\label{teleporation4}
\begin{split}
|0)_x C_{+_1}^{xx}=&~~~~~~~~~~~~~~(C_{+_1}^{xx}+_1C_{-_1}^{xx})|0)_x+_1(A_{+_1}^{xx}+_1A_{-_1}^{xx})|1)_x\\
|1)_x C_{+_1}^{xx}=&~~~~~~~~~~~~~~(A_{+_1}^{xx}-_1A_{-_1}^{xx})|0)_x+_1(C_{+_1}^{xx}-_1C_{-_1}^{xx})|1)_x\\
|0)_y C_{+_1}^{xx}=&|0)_y C_{+_1}^{yy}=(C_{+_1}^{yy}+_1C_{-_1}^{yy})|0)_y+_1(A_{+_1}^{yy}+_1A_{-_1}^{yy})|1)_y\\
|1)_y C_{+_1}^{xx}=&|1)_y C_{+_1}^{yy}=(A_{+_1}^{yy}-_1A_{-_1}^{yy})|0)_y+_1(C_{+_1}^{yy}-_1C_{-_1}^{yy})|1)_y\\
|0)_z C_{+_1}^{xx}=&|0)_z C_{+_2}^{zz}=(C_{+_2}^{zz}+_2C_{-_2}^{zz})|0)_z+_2(A_{+_2}^{zz}+_2A_{-_2}^{zz})|1)_z\\
|1)_z C_{+_1}^{xx}=&|1)_z C_{+_2}^{zz}=(A_{+_2}^{zz}-_2A_{-_2}^{zz})|0)_z+_2(C_{+_2}^{zz}-_2C_{-_2}^{zz})|1)_z\\
|0)_t C_{+_1}^{xx}=&|0)_t C_{+_2}^{tt}=(C_{+_2}^{tt}+_2C_{-_2}^{tt})|0)_t+_2(A_{+_2}^{tt}+_2A_{-_2}^{tt})|1)_t\\
|1)_t C_{+_1}^{xx}=&|1)_t C_{+_2}^{tt}=(A_{+_2}^{tt}-_2A_{-_2}^{tt})|0)_t+_2(C_{+_2}^{tt}-_2C_{-_2}^{tt})|1)_t
\end{split}
\end{equation}

The domain of disjointness (\ref{4Dii}) defines the ``Bell test" $M_{ii}^c$. We use (\ref{4Dii}) to associate each outcome in terms
of different bases in (\ref{teleporation4}) with the corresponding outcomes in terms of the original basis $xx$:

\begin{center}
\begin{tabular}{|l|l|l|l|l|}
\hline
$A_1$&$C_{+_1}^{xx}$&$C_{-_1}^{xx}$&$A_{+_1}^{xx}$&$A_{-_1}^{xx}$\\[0.5ex]
\hline
$|0)_x$&$|0)_x$&$|0)_x$&$|1)_x$&$|1)_x$\\[0.5ex]
\hline
$|1)_x$&$|1)_x$&$|1)_x$&$|0)_x$&$|0)_x$\\[0.5ex]
\hline
$|0)_y$&$|0)_y$&$|1)_y$&$|0)_y$&$|1)_y$\\[0.5ex]
\hline
$|1)_y$&$|1)_y$&$|0)_y$&$|1)_y$&$|0)_y$\\[0.5ex]
\hline
$|0)_z$&$|0)_z$&$|0)_z$&$|1)_z$&$|1)_z$\\[0.5ex]
\hline
$|1)_z$&$|1)_z$&$|1)_z$&$|0)_z$&$|0)_z$\\[0.5ex]
\hline
$|0)_t$&$|0)_t$&$|1)_z$&$|0)_t$&$|1)_t$\\[0.5ex]
\hline
$|1)_t$&$|1)_t$&$|0)_z$&$|1)_t$&$|0)_t$\\[0.5ex]
\hline
\end{tabular}
\end{center}

It is clear that after receiving the result of the test the second party can complete the protocol by implementing local corrections as follows
\begin{equation}
\begin{split}
&C_{+_1}^{xx}\rightarrow I \\
&C_{-_1}^{xx}\rightarrow P_{yt}\\
&A_{+_1}^{xx}\rightarrow P_{xz}\\
&A_{-_1}^{xx}\rightarrow P_{yt}P_{xz},
\end{split}
\end{equation}

The protocol utilized one pair of correlated states and 2 bits of classical communication. Our model resembles quantum mechanics in having the number of classical bits sent larger than the amount of bits of classical information needed to specify an ontic state of $A_1$. In this respect, our model is stronger, than the model of Spekkens, where those two quantities match.\\

Teleportation and Dense Coding are often seen as the manifestation of the characteristic aspects and features of quantum theory and we succeeded in mimicking them in our model. In the future work we will explore additional phenomena and protocols.

\section{Discussion}\label{discussion}

One can commit to the ontic reality of a single physical system in two different ways.

First way is to attribute a set of distinct disjoint physical states to the system. If one follows this route, then in order to mimic quantum
features, e.g. complementarity, non-commutativity, no-cloning and so on, it is essential to introduce a limitation on the access to those states. In other words, one must limit an observer's knowledge about the real state of affairs of the system. When constructing a toy-model one needs to introduce such a limitation ``by hand", as it was done by Spekkens with the help of {\it knowledge balance principle}. In this case all transformations and coherent superpositions are defined for the states of the limited knowledge, e.g. for the epistemic states, not the ontic states. This implies a strongly classical (``ball in a box") approach to ontic states, since it is necessary to rest the rules which govern the transformations of epistemic states on the ontic states of a system. Now, when we are trying to extend the model to include multiple systems and introduce correlations, correlated epistemic states will correspond to probabilistic mixture of product ontic states of individual systems. This picture does not admit a {\it correlated joint ontic states}.

Second way is to relax the requirement from ontic states of a single system and allow them to be non-disjoint. This is the avenue taken in this article. The expectation from all ontic states being mutually disjoint is deeply rooted in our ``classical" experience, but is it not justified as a basic for an ontological commitment. In fact, but putting these classical properties in the very foundation of the toy-model theory we condone this theory to be confined to this classical restriction. Thus, the epistemic picture becomes a self-fulfilling prophecy. The benefit from allowing ontic states of a single system to be disjoint are great. It allows to extend the ontic picture to correlated joint states. In the toy-model suggested in this work, correlated joint states are ontic states. Such states cannot be presented as mixtures of product ontic states of individual systems and therefore are non-separable or entangled, using quantum-mechanical terminology. This model might appear less appealing than models based on the epistemic approach. Indeed, one of the attractive features of Spekkens's model is its elegance and simplicity. It is based on very simple plain assumptions about a single system. Our model does not admit such a simple ``which box the ball is?"-interpretation and therefore is much less intuitive. However, this model demonstrates that if we attempt to maintain the ontic interpretation on the level of composite system, we have to introduce a mathematical structure which resembles quantum theory. We started with very basic assumptions, but in the process of constructing the toy-model we were forced to attribute to it mathematical properties very similar to those that govern quantum theory. One might interpret it as a weakness of the model. Nevertheless, it might be also seen as its strength. In other words, if our basic premiss is that the states of individual systems are ontic, but could be disjoint, and that correlated state are ontic as well, the toy-model theory we get resembles quantum theory in term if its mathematical structure. We see it as an evidence in support of the view that quantum states are ontic states. After all, in nature complementarity, non-commutativity and no-cloning could be inherent features of the states of physical systems on the basic (ontological) level, not the level of access, and this possibility should not be discounted.

\begin{acknowledgments}
I would like to thank Ian Stewart, Tony Short, Shahar Avin for productive and helpful discussions and Jeremy Butterfield for giving me the opportunity to present the ideas of this work at a one-day conference "Symmetry in Physics", Trinity College, Cambridge, March 2011. I am also grateful to Daniel Burgarth for inviting me to Imperial College to present the earlier version of this work.
\end{acknowledgments}


\begin{thebibliography}{30}
\bibitem{comment} Schumacher and Westmoreland have proposed Modal quantum theory \cite{modal_theory}. The nature of their model places it appart from those of Hardy, Spekkens and others. It seems to us that it cannot be used as a support of either ontic or epistemic view.
\end{thebibliography}
\end{document}